\documentclass[lettersize,journal]{IEEEtran}

\usepackage{amsmath,amsfonts}
\usepackage{algorithmic}
\usepackage{array}
\usepackage[caption=false,font=normalsize,labelfont=sf,textfont=sf]{subfig}
\usepackage{textcomp}
\usepackage{stfloats}
\usepackage{url}
\usepackage{verbatim}
\usepackage{graphicx}
\def\BibTeX{{\rm B\kern-.05em{\sc i\kern-.025em b}\kern-.08em
    T\kern-.1667em\lower.7ex\hbox{E}\kern-.125emX}}
\usepackage{balance}

\usepackage{xcolor}
\usepackage{cite} 
\usepackage{hyperref}
\usepackage{multirow}

\newcommand{\E}[1]{\mathbb{E}\left(#1\right)}
\newcommand{\var}[1]{\text{Var}\left(#1\right)}
\newcommand{\quotes}[1]{``#1"}

\newcommand{\nmos}{n_m}
\newcommand{\rmse}{D}
\newcommand{\mse}{\rmse^2}
\begin{document}
\title{Bounds on Agreement between Subjective and Objective Measurements}

\author{Jaden Pieper and Stephen D. Voran, \textit{Senior Member, IEEE}
\thanks{
Received 5 November 2025; revised 3 March 2026; currently under review.

The authors are with the National Telecommunications and Information Administration, Institute for Telecommunication Sciences, Boulder, CO, USA.
}}
\markboth{IEEE Transactions on Multimedia,~Vol.~xx, No.~x, Month~202x}{}

\maketitle

\begin{abstract}
Objective estimators of multimedia quality are typically judged by comparing estimates with subjective ``truth data,’’ most often via Pearson correlation coefficient (PCC) or mean-squared error (MSE).
But subjective test results contain noise, so striving for a PCC of 1.0 or an MSE of 0.0 is neither realistic nor repeatable. 
Numerous efforts have been made to acknowledge and appropriately accommodate subjective test noise in objective-subjective comparisons, typically resulting in new analysis frameworks and figures-of-merit. 
We take a different approach.  By making only the most basic assumptions, we are able to derive bounds on PCC and MSE that can be expected for a subjective test. 

Consistent with intuition, these bounds are functions of subjective vote variance.
When a subjective test includes vote variance information, the calculation of the bounds is straight-forward, and in this case we say the resulting bounds are ``fully data-driven.''
We also provide two options for calculating bounds in cases where vote variance information is not available.
One option is to use vote variance information from other subjective tests that do provide such information, and the second option is to use a model for subjective votes.

Thus we introduce a binomial-based model for subjective votes (BinoVotes) that naturally leads to a mean opinion score (MOS) model, named BinoMOS, with multiple unique desirable properties. 
BinoMOS reproduces the discrete nature of MOS values and its dependence on the number of votes per file. 
This modeling provides vote variance information required by the derived PCC and MSE bounds and
we compare this modeling with data from 18 subjective tests. 
The modeling yields PCC and MSE bounds that agree very well with those found from the data directly. 
These results allow one to set expectations for the PCC and MSE that might be achieved for any subjective test, even those where vote variance information is not available.

\end{abstract}


\section{Introduction}
\IEEEPARstart{A}{ttributes} of multimedia signals such as quality, noisiness, and others are often measured in subjective tests.
In the most common case, subjects watch sets of video files or listen to sets of audio files and rate the overall quality of each file by voting 1 (bad), 2 (poor), 3 (fair), 4 (good), or 5 (excellent).
Mean opinion scores (MOS) are created by averaging responses from multiple subjects for each file.
These tests rely on the underlying assumption that each file has a true quality that can be estimated by running well-designed subjective tests.
Test trials may use a single stimulus or involve comparisons, and may be conducted in labs or performed by crowd-workers.
The International Telecommunication Union has defined specific test protocols for the speech \cite{P.800,P.805,P.806,P.808}, audio \cite{BS.1116}, video \cite{BT.500,P.918,P.915}, multimedia \cite{P.910,P.920}, and gaming  \cite{P.809} domains.

No matter how well a subjective test is designed, the results are imperfect, as all measurements are.
Regardless, these tests are the gold standard for measuring the quality and other attributes of audio, video, or multimedia signals.
They are expensive and time-consuming and this has motivated the 
continual development of objective estimators that can approximate MOS values without querying human subjects.
The development and evaluation of these estimators rely on databases that contain media files and associated subjective test results.
The performance of an objective estimator is often measured by comparing its estimates to the corresponding  subjective test results using mean-squared error (MSE) and Pearson correlation coefficient (PCC). Root mean-squared error (RMSE) is a common alternative to MSE, and we switch between the two here as is convenient.

Results from subjective tests are typically considered \quotes{truth data,} but they are in fact noisy measurements that have inherent uncertainty.
One source of noise in MOS values is the 
discrete nature of subjective rating scales.  Another is the limited number of subjects  that rate each signal.
Additional noise comes from per-subject biases and general test design considerations such as the distribution of quality present in the test, and the level of control in the test environment.
These inherent noises must be acknowledged when developing and assessing the performance of objective estimators.
By accounting for the inherent uncertainty in MOS values we contextualize the degree to which an objective estimator can match this \quotes{truth data.}

We address this need by deriving mathematical bounds for MSE and PCC that are entirely based on the limitations of estimating quality through MOS.
These bounds incorporate the number of votes per file and the distribution of qualities within a test.
The bounds are applicable regardless of how votes, and hence MOS, are generated,
as long as the MOS values converge towards the true quality.
These bounds allow one to decide if an objective estimator has room for improvement, or if it is as good as one could expect given the limitations of the subjective tests being used.

As one might expect, these bounds require information about the variance of the individual votes that contribute to the MOS value.
We call this the vote variance and it drives the ``noise'' that obscures the ``signal.'' 
When vote variance information is available, these bounds let the data \quotes{speak for itself,} without any additional assumptions.
But vote variance is not always shared (this is the case for some of the large crowd-sourced subjective tests commonly used for training objective speech quality estimators).
In this event, a voting model can provide the variance information needed to calculate the bounds.
Thus we also offer a simple and interpretable model for subjective votes.
This model is based on the binomial distribution so we have named it BinoVotes.
Just as real votes are averaged to produce MOS, we average BinoVotes to produce BinoMOS.
We show that BinoVotes and BinoMOS have multiple desirable properties that agree with intuition
and that BinoVotes comport with real votes in the sense that the MSE and PCC bounds for the two are quite close.

Subjective testing has been carefully studied and the popular five-level rating scale and resulting MOS values have been of particular interest.
Their scope and limitations are discussed in detail in \cite{hands2016} which also provides readers with a large number of citations to additional related work.
Seeking to get closer to an underlying truth, researchers have proposed multiple alternatives to simply taking the mean of individual votes.
One proposal involves setting up a maximum likelihood estimation problem and then solving it using belief propagation\cite{li_recover_2017} or alternating projections\cite{li_recover_2020}.
Other proposals include
leveraging generalized linear models\cite{pezzulli_estimation_2021},
an iterative parameter estimation algorithm\cite{tiotsop_scoring_2023},
an approach based on Z-scores\cite{ZREC},
an entropy-based subjective quality recovery algorithm\cite{altieri_subjective_2024},
and a regularized maximum-likelihood estimation algorithm\cite{RMLE}.
Since the underlying truth is never knowable in practice, arguments for and against these alternatives to the mean are necessarily based on various simulations and on robustness to specific classes of controlled perturbations to the votes.

Distributions of votes have previously been modeled with the alpha-stable distribution \cite{gao2023}, the Gaussian distribution \cite{gao2025}, and mixtures of Gaussian distributions\cite{gao2025}. 
The noise inherent in votes and MOS values has been extensively studied and modeled in~\cite{pinson2015} and \cite{pinson2021}.
There has also been significant work on novel  statistics that seek to properly accommodate that noise when judging objective estimators.
Techniques based on classification errors are proposed in \cite{VoranBochum1994, Krasula2016, Ragano2025}, resolving power was introduced in\cite{Brill2002}, and epsilon-insensitive RMSE is described in \cite{P.1401}. 
The approach in \cite{Pinson2023} is highly data-driven.  It uses classification error rates to describe an objective estimator as being equivalent to a subjective test with either 1, 2, 3, 6, or 9 subjects.

These pathways are often valid and useful, but here we offer a unique alternative.
We subscribe to the simple and very defensible axiom that as the number of votes gets large, their mean converges to the true quality of the file contents.
And rather than propose additional specialized figures-of-merit for objective estimators, we derive bounds for two existing, commonly used, principled performance measures: PCC and MSE. 

An additional distinction of our work is that no model for MOS uncertainty is needed.  This is very desirable because such models can be problematic --- they can violate the discrete nature of MOS values or the finite range of MOS values, and ad-hoc fixes like clipping can introduce additional complications that include new biases.
These are not issues in our work because it is firmly grounded in the true mathematical properties of MOS, including its discrete nature and the fact that the set of allowable discrete MOS values is a function of the number of votes used for a MOS value. From this basis we derive expected values for PCC and MSE.   We are not aware of any previous work that bounds PCC or MSE by using only the intrinsic properties of votes and MOS.

The next section presents the properties of rating scales and the properties that MOS inherits from those scales.
In Section \ref{sec:performance-bounds} we leverage those properties to derive bounds for PCC and MSE.
We present a new binomial distribution-based voting model, BinoVotes, in Section \ref{sec:binovotes} and its natural by-product, BinoMOS, in Section \ref{sec:binoMOS}.
In Section~\ref{sec:binomos-bounds} we derive bounds for PCC and MSE under a BinoVotes voting model.
In Section~\ref{sec:application} we present three approaches for estimating bounds from  real subjective test results.
Finally, in Section \ref{sec:data-analysis} we examine results from 22 different subjective tests. 
Eighteen of these include vote variance information that allows us to create fully data-driven PCC and MSE bounds, and then compare them with the bounds that result from assuming the BinoVotes voting model as well as using an empirically-based vote variance assumption. 
The remaining four test results do not include vote variance information and we calculate PCC and MSE bounds in two different ways. 

\section{Mathematical Properties of Rating Scales and MOS}
We now consider the rating of an arbitrary media signal. 
There are many different properties that can be rated, but the most common is ``overall quality'' and all of the subjective test results in Section \ref{sec:data-analysis} pertain to ``overall quality.'' 
For conciseness we use the term ``quality'' in the following but all mathematical results are independent of the property rated. 
The comparisons in Section \ref{sec:data-analysis} could be extended to the rating of other properties in future research efforts.

We denote the true quality of the contents of a media file as $Y.$
We define some continuous quality scale such that $Y\in[s_L, s_H]$.
To assess $Y$ in a subjective test, we define a rating scale as
\begin{equation}\label{eqn:general-rating-scale}
S = \left\{\left(\frac{s_H - s_L}{n_s - 1}\right ) k + s_L\right\}_{k=0}^{n_s - 1}.
\end{equation}
The rating scale is a set of $n_s$ discrete values that span the continuous quality scale. 
In a subjective test, each subject rates the quality of a given media file by selecting or ``voting for'' one of these $n_s$ values according to their perception and opinion of the media file. 
Thus $n_s$ defines the resolution of an individual vote. 
Note that the special case $s_L\,=\,1$, $s_H\,=\,5$, and ${n_s\,=\,5}$ gives the very popular five-point subjective testing scale, sometimes called ``the MOS scale.'' 
Quality exists on a continuous scale and subjective tests produce sets of votes that correspond to $n_s$ discrete points on that scale.

In a subjective test, $n_v$ independent voters or subjects assess the quality of a media file.
We denote these votes, or ratings, as $\{R_j\}_{j=1}^{n_v}$, where $R_j \in S$.
The MOS of a media file is defined as
\begin{align}\label{eqn:mos-def}
    X = \frac{1}{n_v} \sum_{j=1}^{n_v} R_j.
\end{align}
It follows from (\ref{eqn:general-rating-scale}) and (\ref{eqn:mos-def}) that the domain of the MOS values is
\begin{equation}
M = \left\{\left(\frac{s_H - s_L}{n_v (n_s - 1)}\right )k + s_L\right\}_{k=0}^{n_v(n_s - 1)},
\end{equation}
so that $X \in M$.
This result makes it clear that, given a fixed rating scale $S$, as the number of votes per media file, $n_v$, increases, the number of possible MOS values and resolution of the MOS values both  increase. 
So one of the factors that limits the ability of MOS to agree with true quality $Y$ is the number of votes per file.
For example, on the commonly used 1 to 5 quality scale, if a file has a quality of 3.30 and only one vote, then the smallest possible error would be 0.30. 
But with two votes the smallest possible error decreases to 0.20, and with three votes that bound is approximately 0.03.

In this way, for a fixed rating scale $S$, the number of votes per file is the most obvious contributor to differences between MOS and true quality.
Additional differences come from the random behavior of subjects and how votes tend to be distributed across the quality range.
If the true quality is 3.3, subjects do not only vote that the quality is 3 or 4 in the proper proportions.
Instead, in general, they use the entire range of the quality scale.
This increases 
the vote variance and hence the 
MOS variance, and can only really be alleviated by increasing the number of votes per file.
Increasing the number of votes per file can be thought of in terms of the central limit theorem; it is essentially decreasing the variance of the distribution of the MOS.

We model the individual votes, $R_j$, tendered by a group of subjects for a single file as independent but identically distributed (i.i.d.) random variables. 
If one conditions this vote distribution by subject, 
the conditional distributions may depend on the subject,
which is often called ``per-subject bias."
This is completely consistent with modeling the set of \emph{all votes from all subjects} for a given file as i.i.d. random variables from a single distribution which is independent of individual subjects.

We say that a mathematical model for votes is  ``well-behaved'' if given a true quality, the expected value of the votes is that quality,
\begin{align}\label{eqn:good-vote}
    \E{R_j|Y} = Y.
\end{align} 
In this work we assume that votes come from a well-behaved model and in fact this assumption is the bedrock of all subjective tests. 

If votes are well-behaved, then MOS is also well-behaved. 
In fact, MOS is well-behaved if and only if votes are well-behaved, which is demonstrated at the end of this section.
The votes $R_j$ are random variables and  their mean $X$ is also a random variable. 
Using (\ref{eqn:mos-def}) we see that the MOS, $X$, given a true quality $Y$, also has a mean of the true quality
\begin{align}\label{eqn:good-mos}
    \E{X|Y} = Y.
\end{align}
Note that the law of total expectation further shows
\begin{align}
    \E{R_j} &= \E{\E{R_j|Y}} = \E{Y}, \label{eqn:vote-expected-value} \\
    \E{X} &= \E{\E{X|Y}} = \E{Y}. \label{eqn:mos-expected-value}
\end{align}

The votes $R_j$ also have some variance.  
The rating scale is finite and the only way to satisfy (\ref{eqn:good-mos}) at either end of the scale is to have zero variance there.  
But variance cannot be zero across the entire rating scale.
This motivates us to define the conditional vote variance function $v_r(Y)$ as
\begin{align}\label{eqn:conditional-vote-variance}
    v_r(Y) = \var{R_j|Y}, \;\;\; \forall j = 1,\cdots,n_v .
\end{align}
Then the variance of MOS across a subjective test can be found using the law of total variance and (\ref{eqn:good-mos}),
\begin{align}\label{eqn:mos-variance-1}
    \var{X} 
    &= \E{\var{X|Y}} + \var{Y}.
\end{align}
Using (\ref{eqn:mos-def}), (\ref{eqn:conditional-vote-variance}), and the fact that ratings $R_j$ are independent, we can show that the conditional variance of the MOS is
\begin{align}\label{eqn:conditional-mos-variance}
    \var{X|Y} 
    &= \frac{1}{n_v^2} \sum_{j=1}^{n_v} \var{R_j|Y} 
    = \frac{v_r(Y)}{n_v}.
\end{align}
We can plug (\ref{eqn:conditional-mos-variance}) into (\ref{eqn:mos-variance-1}) to find that the variance of MOS is
\begin{align}\label{eqn:mos-variance}
    \var{X} 
    &= \frac{\E{v_r(Y)}}{n_v} + \var{Y}.
\end{align}
This result shows that the variance in MOS is the variance of the true quality distribution, $\var{Y}$, increased by a term that captures the vote variance.
This additional variance is scaled down as the number of votes increases, so that regardless of how much variance the voting adds, gathering enough independent votes will cause MOS values to converge to true quality values and MOS variance to converge to the variance of the true quality distribution.

As mentioned above, subjects may exhibit individualized voting behaviors or biases.
This is fully compatible with our modeling of votes and MOS even though our modeling so far has not required any explicit bias terms.
We conclude this section by 
directly addressing per-subject bias within the vote model we have developed.

Those who conduct subjective tests agree that MOS values are well-behaved --- by collecting enough votes from voters with different biases, MOS will converge to the true quality. 
In fact, MOS is well-behaved if and only if votes are well-behaved.
This can be shown by observing the following.
Per-subject bias can be described as a subject $V_j$ giving votes that have a biased expected value, as 
\begin{align}
\label{eqn:biasedVotes}
\E{R_j|Y, V_j} = Y + \delta_j,
\end{align}
where $\delta_j$ is an i.i.d. random variable.
Applying the law of total expectation to (\ref{eqn:biasedVotes}) gives
\begin{align}\label{eqn:expected-vote-with-bias}
    \E{R_j|Y} = \E{\E{R_j | Y, V_j}} = Y + \E{\delta_j}.
\end{align}
The law of total expectation also shows that the expected value of MOS derived from votes with individual i.i.d. biases is
\begin{align}\label{eqn:expected-mos-with-bias}
    \E{X|Y}  
    &= Y + \E{\delta}.
\end{align}
If we assume MOS is well-behaved, then (\ref{eqn:expected-mos-with-bias}) implies that $\E{\delta}=0$  and (\ref{eqn:expected-vote-with-bias}) simplifies to (\ref{eqn:good-vote}).
This means that a vote model with per-subject biases that produces well-behaved MOS values also produces well-behaved votes.
If we instead assume that votes are well-behaved, then (\ref{eqn:expected-vote-with-bias}) implies that $\E{\delta_j}=0,\;\;\forall j$, and we can see from (\ref{eqn:expected-mos-with-bias}) that MOS must be well-behaved.
In this work, the only relevant impact of per-subject biases is that they increase the conditional vote variance, $v_r(Y)$.


\section{Performance Bounds for Objective Estimators of MOS}\label{sec:performance-bounds}
In order to bound the performance of objective estimators, we consider the best possible estimator.
This best-case objective estimator would be the oracle that can access the true quality $Y$.
This means that agreement between true quality $Y$ and MOS $X$ bounds the agreement between any objective estimator and MOS. 
Specifically, the PCC between $Y$ and $X$ gives an upper bound for the PCC that a real objective estimator can achieve. 
And the MSE between $Y$ and $X$ gives a lower bound for the MSE that a real objective estimator can achieve.
\footnote{Estimators are sometimes followed by a non-linear mapping designed to improve agreement with MOS. 
Such mappings do not conflict with the use of the bounds we derive here.
The bounds are derived by considering the oracle estimator which produces estimates that are, in fact, the exact true quality. 
No mapping is required for the oracle estimator, and the bounds that follow from it are directly related to the noise in the MOS values.}

We now consider a subjective test with a set of $n_f$ media files, $\{F_i\}_{i=1}^{n_f}$, with associated true qualities $\{Y_i\}_{i=1}^{n_f}$, such that the quality of file $F_i$ is $Y_i$.
Let $Y_i$ be i.i.d. according to some arbitrary, continuous distribution $f_Y$ over $[s_L, s_H]$.
Each file $F_i$ is rated by $n_v$ subjects that give well-behaved votes, so MOS is well-behaved and (\ref{eqn:good-mos}) is satisfied. 
The MOS values $X_i$ are i.i.d. and $X_i$ is our best available measurement of $Y_i$.
In actual subjective tests, there is no access to the hidden variable $Y_i$ that describes the true quality.  The only way to get information about it is through votes and the MOS values they produce, $X_i$.

\subsection{MSE Lower Bound}

The squared error between $X_i$ and $Y_i$ is the random variable
\begin{align}\label{eqn:squared-error-def}
    \epsilon_i^2 = (X_i - Y_i)^2.
\end{align}
The mean-squared error is also a random variable and is the mean of these squared errors,
\begin{align}\label{eqn:mse-def}
    \mse = \frac{1}{n_f}\sum_{i=1}^{n_f} \epsilon_i^2.
\end{align}
The expected value of the MSE is
\begin{align}\label{expected-mse}
    \E{\mse} = \frac{1}{n_f} \sum_{i=1}^{n_f} \E{\epsilon_i^2}
    &= \E{\epsilon^2},
\end{align}
where the final equality exploits the fact the $\epsilon_i^2$ are i.i.d.
Using the law of total expectation we can show
\begin{align}\label{eqn:expected-squared-error}
    \E{\epsilon^2} &= \E{(X - Y)^2}
    = \E{\E{(X - Y)^2 | Y}} \nonumber \\
    &= \E{\E{(X - \E{X})^2|Y}}
    = \E{\var{X|Y}}. 
\end{align}
Using (\ref{eqn:conditional-mos-variance}), (\ref{eqn:expected-squared-error}), and (\ref{expected-mse}) we find that the expected value of the MSE is
\begin{align}\label{eqn:mse-bound}
    \E{\mse} = \E{\epsilon^2} &= \frac{\E{v_r(Y)}}{n_v}.
\end{align}

This is the expected value of the MSE between MOS and true quality. 
Following the exposition at the start of this section, this is also the lower bound for the expected value of MSE between MOS and any objective estimator of quality.

The result at (\ref{eqn:mos-variance}) provides an alternative interpretation of the MSE result in (\ref{eqn:mse-bound}):  $\E{\mse} = \var{X} - \var{Y}$.  
This has a very natural interpretation: the expected MSE is the additional variance that subjective test results (MOS values) impose on the true quality distribution.

\subsection{Pearson Correlation Coefficient Upper Bound for a Population}
We now derive a population-based Pearson correlation coefficient bound. 
Of course every subjective test has a finite number of samples so a bound on the sample correlation may seem more appropriate.
However, the expected value of the sample correlation converges rapidly to the population correlation and this is demonstrated in Section \ref{ssec:samplecorrelations}.

For a population, the PCC is defined as
\begin{align}\label{pccDef}
    \rho(X, Y) = \frac{\E{XY} - \E{X}\E{Y}}{\sqrt{\var{X}\var{Y}}}.
\end{align}
By using the law of total expectation and (\ref{eqn:good-mos}) we can see that
\begin{align}\label{eqn:exy}
    \E{XY} \!=\! \E{\E{XY|Y}}
    = \E{Y\E{X|Y}}    = \E{Y^2}.
\end{align}
From (\ref{eqn:mos-expected-value}) we know that $\E{X} = \E{Y}$.  Using this fact and (\ref{eqn:exy}) in (\ref{pccDef}) gives
\begin{align}\label{eqn:corr-bound-1}
    \rho(X, Y) 
    &= \frac{\var{Y}}{\sqrt{\var{X}\var{Y}}}
    = \sqrt{\frac{\var{Y}}{\var{X}}}.
\end{align}
Finally, by plugging in (\ref{eqn:mos-variance}) we see
\begin{align}
    \rho(X,Y) &= \sqrt{\frac{\var{Y}}{\var{Y} + \frac{1}{n_v} \E{v_r(Y)}}} \label{eqn:corr-bound} \\
    &= \sqrt{\frac{\var{X} - \frac{1}{n_v}\E{v_r(Y)}}{\var{X}}} \label{eqn:corr-bound-mos}.
\end{align}
Equation (\ref{eqn:corr-bound}) is useful when working from assumed distributions of true quality, $Y$, as in Sections~\ref{sec:binovotes}, ~\ref{sec:binoMOS}, and \ref{sec:binomos-bounds}. Equation (\ref{eqn:corr-bound-mos}) is useful when working from subjective test results, which yield MOS distributions, $X$, as in Sections~\ref{sec:application} and \ref{sec:data-analysis}.
In practice, $Y$ is always a hidden random variable, however, (\ref{eqn:mos-expected-value}) and (\ref{eqn:mos-variance}) allow us to estimate its first and second moments from MOS distributions and voting variance information.
Alternately, in terms of MSE, these become
\begin{align}\label{eqn:corr-bound2}
    \mkern-11mu \rho(X,Y)\! =\!\sqrt{\frac{\var{Y}}{\var{Y} + \E{\mse}}} 
    \!= \sqrt{\frac{\var{X} - \E{D^2}}{\var{X}}},\!
\end{align}
where these expressions emphasize that as the MSE goes to zero the PCC goes to one. 
Equation (\ref{eqn:corr-bound2}) gives the population correlation between MOS and true quality. Following the exposition at the start of this section, this is also the upper bound for the correlation between MOS and any objective estimator of quality.\footnote{
Spearman's rank correlation coefficient (SRCC) is often considered alongside of PCC.
Future work could involve extending these bounds to include alternative agreement statistics such as SRCC.
}

We now have bounds for the MSE (\ref{eqn:mse-bound}) and the PCC (\ref{eqn:corr-bound2}). The MSE bound requires only the expected variance of individual votes and the number of votes.  The PCC bound requires the MSE bound and either the variance of the MOS values $X$ or the variance of the underlying true quality distribution $Y$.
These bounds provide information on the performance limitations inherent in the subjective tests.
In other words, they provide context for how the discrete nature of MOS limits the MSE and PCC values that can be achieved for different subjective tests.
When an objective estimator produces results near the bounds for a given subjective test, there is likely little room for true improvement left, even if the MSE is not near zero or the correlation is not near one.

\section{A Binomial Model for Votes --- BinoVotes}
\label{sec:binovotes}
We now introduce a simple, intuitive, and effective model for subjective votes.
The model honors the finite and discrete natures of the rating scale, and thus leads to MOS values that also respect these attributes.
We argue that this is far preferable to approximating votes or MOS with continuous random variables.

We model individual votes, or ratings, via the binomial distribution, which is a natural fit for this problem.
It is a discrete distribution with a finite range and is the discrete analog of the normal distribution.
In light of the binomial heritage of our voting model, we named it ``BinoVotes.''

The binomial distribution models the probability of a given number of successful Bernoulli trials with success parameter $p$ in $n$ independent trials.
For a random variable \mbox{$B\sim\text{Binomial}(n, p)$,} the probability mass function (PMF) is
\begin{equation}
    P(B = k) = {n\choose k} p^k (1 - p)^{(n - k)}, \;\;\; k=0, 1, \cdots, n.
\end{equation}
The expected value and variance, respectively, are 
\begin{align}
    &\E{B} = np,\label{eqn:binomial-expected-value} \\
    &\var{B} = np(1-p).\label{eqn:binomial-variance}
\end{align}

We translate the binomial distribution domain of support to match an arbitrary rating scale as given in (\ref{eqn:general-rating-scale}) by setting the number of trials $n$ to be one less than the number of discrete levels in the rating scale, i.e., $n = n_s-1$.
The BinoVotes random variable for a single subject, $j$, is 
\begin{align}\label{eqn:binovote}
    R_j = \frac{s_H - s_L}{n_s - 1}B_j + s_L, \;\; j=1,\hdots, n_v,
\end{align}
where the $B_j$ is the binomial random variable
\begin{align}\label{eqn:binovotes-binomial}
    B_j \sim \text{Binomial}\left((n_s - 1) , \frac{Y - s_L}{s_H - s_L}\right).
\end{align}
Here $Y\in[s_L, s_H]$ is a random variable that represents the true quality, and $Y$ is drawn from an arbitrary distribution, $f_Y$.
We say that $R_j \sim \text{BinoVotes}(n_s, Y)$.
Note that the domain of support for $R_j$ is the rating scale $S$, defined in (\ref{eqn:general-rating-scale}).

The BinoVotes random variable $R_j$ satisfies
the condition for a well-behaved vote model given in (\ref{eqn:good-vote}). This is demonstrated by using the expected value in (\ref{eqn:binomial-expected-value}) and the definitions in (\ref{eqn:binovote}) and (\ref{eqn:binovotes-binomial}):
\begin{align}
    \E{R_j|Y} 
    &= s_L + \frac{(s_H - s_L)(n_s - 1)(Y - s_L)}{(n_s - 1)(s_H - s_L)} 
    &= Y.
\end{align}
The conditional variance of the vote $R_j$ is found by using (\ref{eqn:binomial-variance}) with the definitions in (\ref{eqn:binovote}) and (\ref{eqn:binovotes-binomial}):
\begin{align}
    \var{R_j| Y} 
    &= \left(\frac{s_H - s_L}{n_s - 1}\right)^2 \frac{(n_s - 1)(Y - s_L)(s_H - Y)}{(s_H - s_L)^2} \nonumber \\
    &= \frac{(Y- s_L)(s_H - Y)}{n_s - 1}.
\end{align}
So for BinoVotes, the function $v_r(Y)$ that gives vote variance as a function of true quality is simply
\begin{align}\label{eqn:binovotes-variance}
    v_r(Y) = \frac{(Y- s_L)(s_H - Y)}{n_s - 1}.
\end{align}
From Section \ref{sec:performance-bounds} we know that the expected value of this function will be of great utility in what follows, 
\begin{align}\label{eqn:bv-expected-vote-varianccc}
    \E{v_r(Y)} 
    &= \frac{1}{n_s - 1}\left((s_H + s_L)\E{Y} -\E{Y^2} - s_Ls_H \right) \nonumber \\
    &= \frac{(\E{Y} - s_L)(s_H - \E{Y}) - \var{Y}}{n_s - 1}.
\end{align}

Once a rating scale is selected, BinoVotes models votes using a single parameter $Y$.
At first glance, having only a single parameter may seem like a disadvantage.
But the binomial approach is actually very natural because that single parameter links the variance of the votes to the mean of the votes in the appropriate way.
Specifically, as (\ref{eqn:binovotes-variance}) shows, the variance is zero at both ends of the quality scale, as it must be.  And at the center of the scale, where the diversity of perceptions and diversity of votes is likely to be the largest, the variance takes its maximum value
This has previously been discussed in the context of $s_L=1$, $s_H=5$ and we relate our current BinoVotes results to this aspect of previous work in Sec.~\ref{sec:common-mos}.

\begin{figure}[t]
    \centering
    \includegraphics[width=\linewidth]{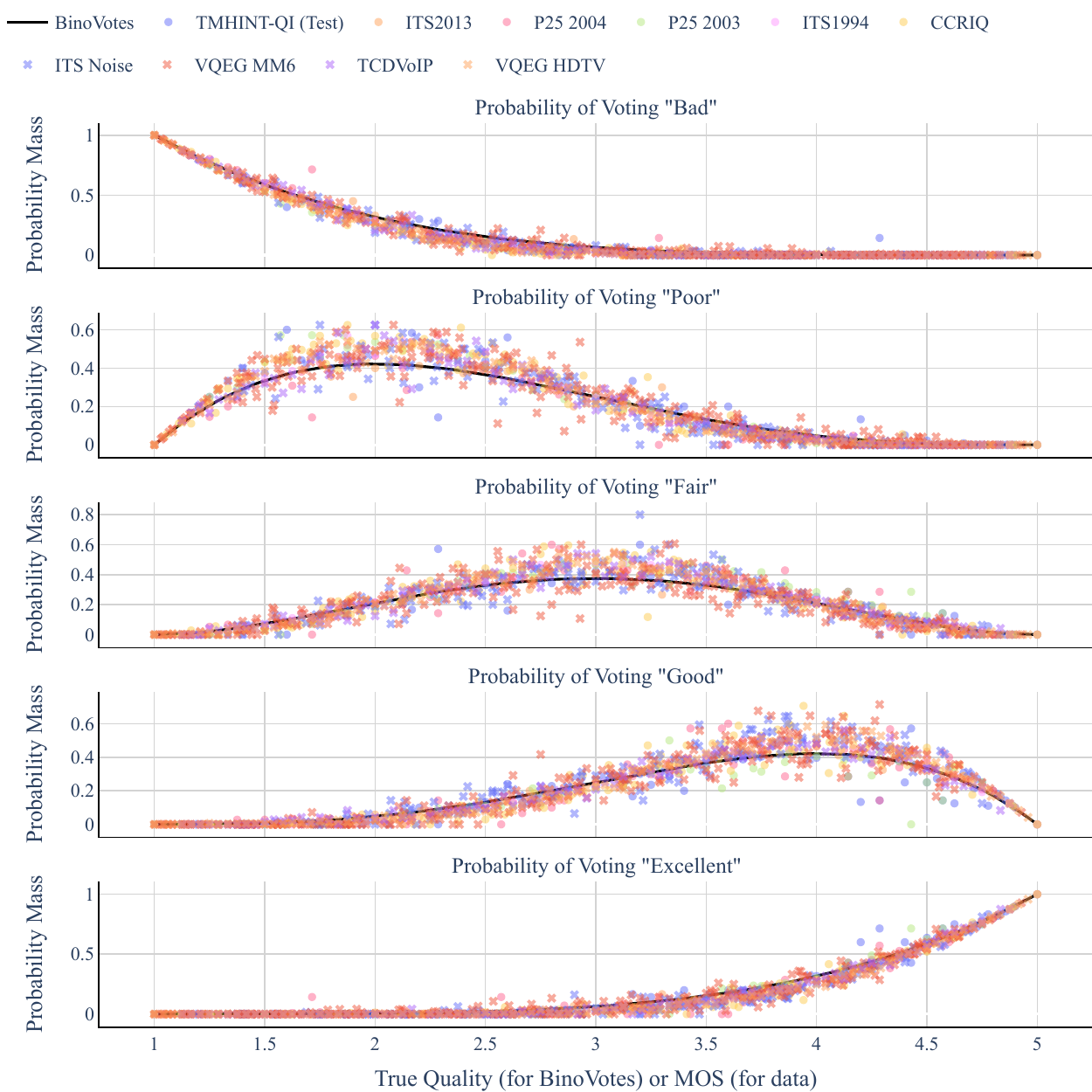}
   \caption{Probabilities for the five ratings of the common MOS scale for the BinoVotes model (vs. true quality $Y$) and data (vs. MOS) for 10 tests. \vspace{0mm}}
    \label{fig:vote-pmf}
\end{figure}
Figure~\ref{fig:vote-pmf} shows BinoVotes PMFs (as functions of $Y$) along with vote distributions (as functions of MOS) for 10 subjective tests that include individual votes, which will be introduced in more detail in Section~\ref{sec:data-analysis}.
For each test, for a given MOS value the fraction of votes that are labeled ``Bad,'' ``Poor,'' ``Fair,'' ``Good,'' or ``Excellent'' are plotted on their respective graphs.
This figure demonstrates that, while not perfectly accurate, BinoVotes does capture the correct trends for how subjects vote and that the binomial distribution is a very good basis for a voting model.

\section{MOS from BinoVotes --- BinoMOS}\label{sec:binoMOS}
Following standard practice, we average BinoVotes to produce BinoMOS.  That is, when $R_j \sim \text{BinoVotes}(n_s, Y)$ and $X$ is
\begin{align}
    X = \frac{1}{n_v} \sum_{j=1}^{n_v} R_j,
\end{align}
We say that $X \sim \text{BinoMOS}(n_v, Y)$. 
The distribution also depends on $s_L$, $s_H$, and $n_s$, which are used implicitly in the definition of $S$ and the domain of support for $Y$.

Since BinoVotes satisfies (\ref{eqn:good-vote}) and (\ref{eqn:conditional-vote-variance}), it follows that the BinoMOS random variables satisfies $\E{X|Y} = Y$ (\ref{eqn:good-mos}) and that $\E{X} = \E{Y}$ (\ref{eqn:mos-expected-value}) .
Further we can use (\ref{eqn:mos-variance}) and (\ref{eqn:bv-expected-vote-varianccc}) to show that the BinoMOS variance is 
\begin{align}\label{binMOSvariance}
    \!\!\!\var{X}\!
    =\! \frac{(\E{Y} \!-\! s_L)(s_H \!-\! \E{Y}) \!+\! (n_m\! -\! 1)\var{Y}}{n_m}, 
\end{align}
where $n_m = n_v (n_s - 1)$. Note that the total number of distinct values that $X$ can take is $n_m + 1$.
Increasing the number of votes per file, $n_v$, will increase $n_m$ and will decrease the BinoMOS variance, as desired.

Next we find the PMF for BinoMOS.
We will leverage the fact that the sum of independent binomial random variables is a binomial random variable.  Specifically, when $n_v$ independent $\text{Binomial}(n_s-1, p)$ random variables are summed, the resulting random variable is $\text{Binomial}(n_v(n_s-1), p)$ and this random variable describes $X$, 
up to the same scale and shift factors seen in (\ref{eqn:binovote}).
Since $X \sim \text{BinoMOS}(Y, n_v)$, the $n_m+1$ allowable values of $X$ are $x_k = s_L + k\frac{s_H - s_L}{\nmos}$, \mbox{$k=0, 1, \cdots, n_m$.}
It follows that the conditional probability of each of these allowable values is given by 
\begin{align}
    P(X = x_k|Y=y) = P(B = k),
\end{align}
where $B\sim \text{Binomial}(\nmos, \frac{y - s_L}{s_H - s_L})$.
And the p.m.f of X simplifies to
\begin{align}\label{eqn:binovotesPMF}
    &P(X = x_k) = \int_{s_L}^{s_H} P(X=x_k|Y=y) f_Y(y)dy \nonumber \\
    &= {\nmos \choose k}\frac{ \int_{s_L}^{s_H} (y - s_L)^k (s_H - y)^{\nmos - k}f_Y(y) dy}{(s_H - s_L)^{\nmos}}.
\end{align}
Note that for the special case where $Y$ follows a scaled and shifted beta distribution, the BinoMOS distribution is described by a scaled and shifted beta-binomial distribution.

\section{Performance Bounds for Objective Estimators of BinoMOS}
\label{sec:binomos-bounds}

We can now use the BinoVotes and BinoMOS results from Sections \ref{sec:binovotes} and \ref{sec:binoMOS} in the general performance bounds of Section \ref{sec:performance-bounds} to find performance bounds for objective estimators in the case where subjective test votes are consistent with the BinoVotes model.
The MSE lower bound is found by using the expected BinoVote variance (\ref{eqn:bv-expected-vote-varianccc}) in the general MSE bound~(\ref{eqn:mse-bound}):
\begin{align}\label{eqn:mse-bound-binovotes}
    \E{\mse} 
    &= \frac{(\E{Y} - s_L)(s_H - \E{Y}) - \var{Y}}{n_m}.
\end{align}

Similarly, the population PCC upper bound is found by using BinoMOS MSE (\ref{eqn:mse-bound-binovotes}) in the general PCC bound (\ref{eqn:corr-bound2}):
\begin{align}\label{eqn:corr-pop-bound}
    &\rho(X, Y)  = \sqrt{\frac{\var{Y}}{\var{Y}+\E{\mse}}}
    \nonumber \\
    &= \sqrt{\frac{\nmos \var{Y}}{(\nmos-1)\var{Y} + (\E{Y} - s_L)(s_H - \E{Y}) }}\,.
\end{align}
Having assumed the BinoVotes model for votes, we now have bounds for MSE and the PCC
that are completely defined in terms of the expected value and variance of the underlying true quality distribution $Y$. 

\subsection{BinoVotes and BinoMOS Examples on the Common MOS Scale}\label{sec:common-mos}

We now offer some specific BinoVotes and BinoMOS examples for the very common 1 to 5 MOS scale, that is, $s_L = 1$, $s_H = 5$, and $n_s=5$. 
In this case (\ref{eqn:binovote}) and (\ref{eqn:binovotes-binomial}), respectively, become
\begin{align}
    R_j &= B_j + 1, \\
    B_j &\sim\text{Binomial}\left(4, \frac{Y - 1}{4}\right).
\end{align}
The vote variance function (\ref{eqn:binovotes-variance}) becomes
\begin{align}\label{eqn:binovotes-1-5-vote-var}
    v_r(Y) = \frac{(Y - 1)(5 - Y)}{4},
\end{align}
and its expected value (\ref{eqn:bv-expected-vote-varianccc}) is
\begin{align}\label{eqn:binovotes-vote-variance}
    \E{v_r(Y)} = \frac{(\E{Y} - 1)(5 - \E{Y}) - \var{Y}}{4}.
\end{align}
The MSE lower bound (\ref{eqn:mse-bound-binovotes}) becomes
\begin{align}
\label{eqn:BVMSEbound}
    \E{\epsilon^2} = \frac{(\E{Y} - 1)(5 - \E{Y}) - \var{Y}}{4 n_v},
\end{align}
and the PCC upper bound (\ref{eqn:corr-pop-bound}) becomes
\begin{align}
    \label{eqn:BVPCCbound}
    \rho(X, Y) = \sqrt{\frac{4n_v \var{Y}}{(4n_v - 1) \var{Y} + (\E{Y} - 1)(5 - \E{Y})}}.
\end{align}

\begin{figure}[!t]
\centering
\includegraphics[width=\linewidth]{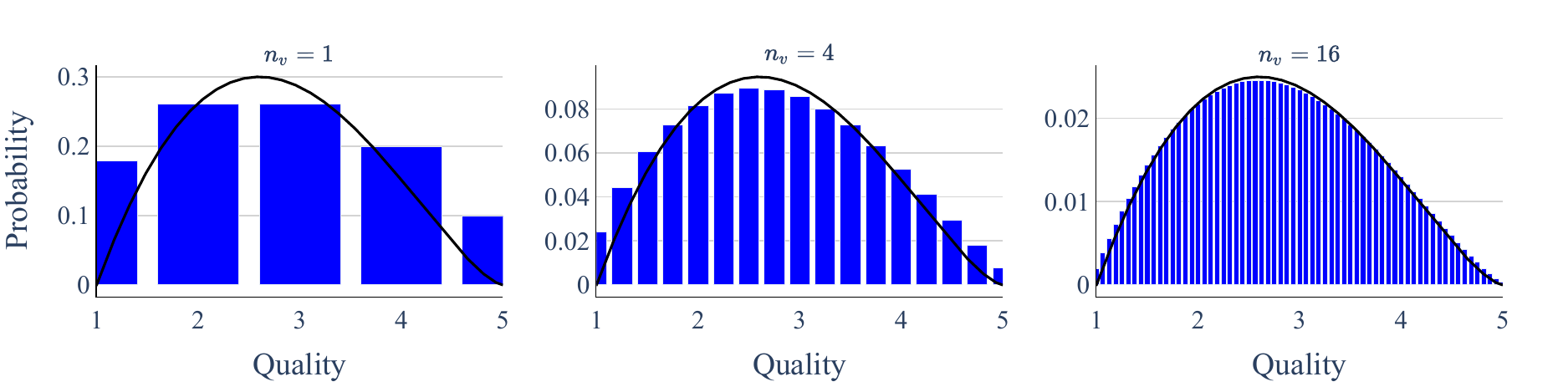}
\label{fig_second_case}
\caption{
Example BinoMOS PMFs  (blue) when true quality distribution is $\text{Beta}(2, 2.5)$ (black, scaled for visual comparison). 
Number of votes per file, $n_v$, increases from left to right with the values 1, 4, and 16. 
\vspace{0mm}
}
\label{fig:example-beta}
\end{figure}

We use the BinoMOS PMF result in (\ref{eqn:binovotesPMF}) to plot example PMFs in Figure \ref{fig:example-beta}.
The figure shows PMFs for three values of $n_v$ when $f_Y$ is Beta(2, 2.5).
Consistent with real votes, the BinoMOS PMF is a discrete approximation of the underlying  continuous distribution of true quality. 
As the number of votes per file increases, the resolution of the BinoMOS values also increases and the approximation becomes closer.

\begin{figure}[h]
    \centering
    \subfloat[]{
        \includegraphics[width=0.49\linewidth]{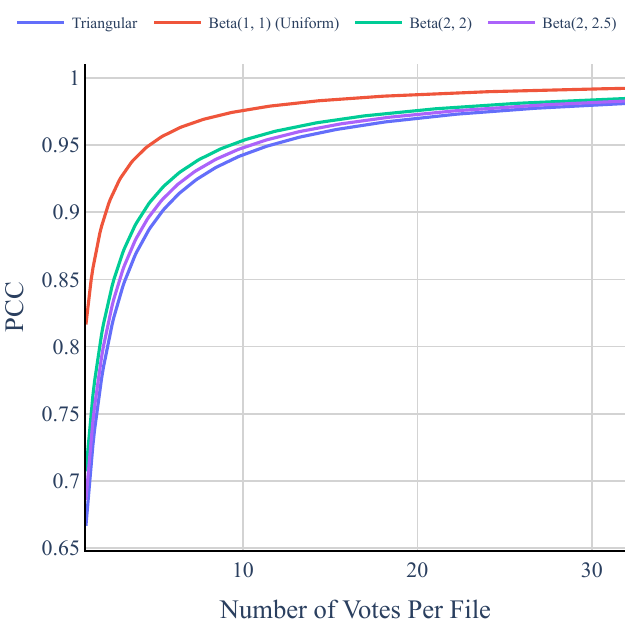}
        \label{fig:general-corr-bounds}
        }
    \subfloat[]{
        \includegraphics[width=0.49\linewidth]{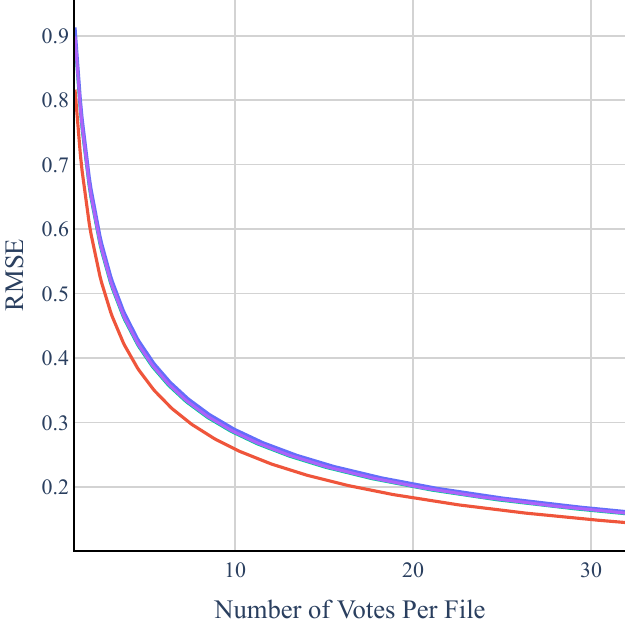}
        \label{fig:general-rmse-bounds}
        }
    \caption{
    Agreement statistic bounds versus number of votes per file, $n_v$, for four example true quality distributions, $f_Y(y)$.
    (a) PCC upper bounds. 
    (b) RMSE lower bounds. The Triangular, Beta(2, 2), and Beta(2, 2.5) bound lines are visually indistinguishable. 
    \vspace{0mm}
    }
    \label{fig:general-bounds}
\end{figure}
Figure~\ref{fig:general-bounds} shows PCC and RMSE bounds for a variety of distributions of $f_Y$ as a function of the number of votes per file.
The plots make it clear that when more votes per file are available it is possible to achieve lower RMSE and higher PCC between the mean of those votes and an objective estimator.
The underlying quality distribution is less important, and bounds for the three non-uniform distributions are close to each other for PCC and nearly identical for RMSE.

The parabola $(Y-1)(5-Y)$ in (\ref{eqn:binovotes-1-5-vote-var}) shapes the vote variance across the quality range and is a direct mathematical consequence of the BinoVotes model. 
In \cite{Hossfeld2011} a derivation of the maximum possible vote variance across the quality range produces the same parabola.  
This is followed by fitting scaled versions of the parabola to 28 sets of subjective test results that cover a wide range of application areas (see Table 1 in \cite{Hossfeld2011}).
The median value of the 28 scale factors is 0.24, which is quite close to the scale factor of $\frac{1}{4} = 0.25$ seen in (\ref{eqn:binovotes-1-5-vote-var}).
That is, the BinoVotes variance function is ``in the middle of'' the 28 variance versus MOS relationships studied in \cite{Hossfeld2011}.
The image quality assessment work in \cite{gao2025} requires that the parabola be fitted to six sets of subjective image quality test results. The resulting median value is 0.15 which is somewhat lower than the scale factor found in the BinoVotes vote variance function.
In \cite{tiotsop_scoring_2023} a scaled version of that same parabola is introduced as an ``inconsistency factor''  to ``model the fact that we expect larger inconsistency at the center of the scale, while going towards the extremes, it must decrease progressively towards zero.'' The BinoVotes model achieves this innately.

\subsection{Pearson Correlation for a Sample}
\label{ssec:samplecorrelations}
\begin{figure}[t]
    \centering
    \includegraphics[width=\linewidth]{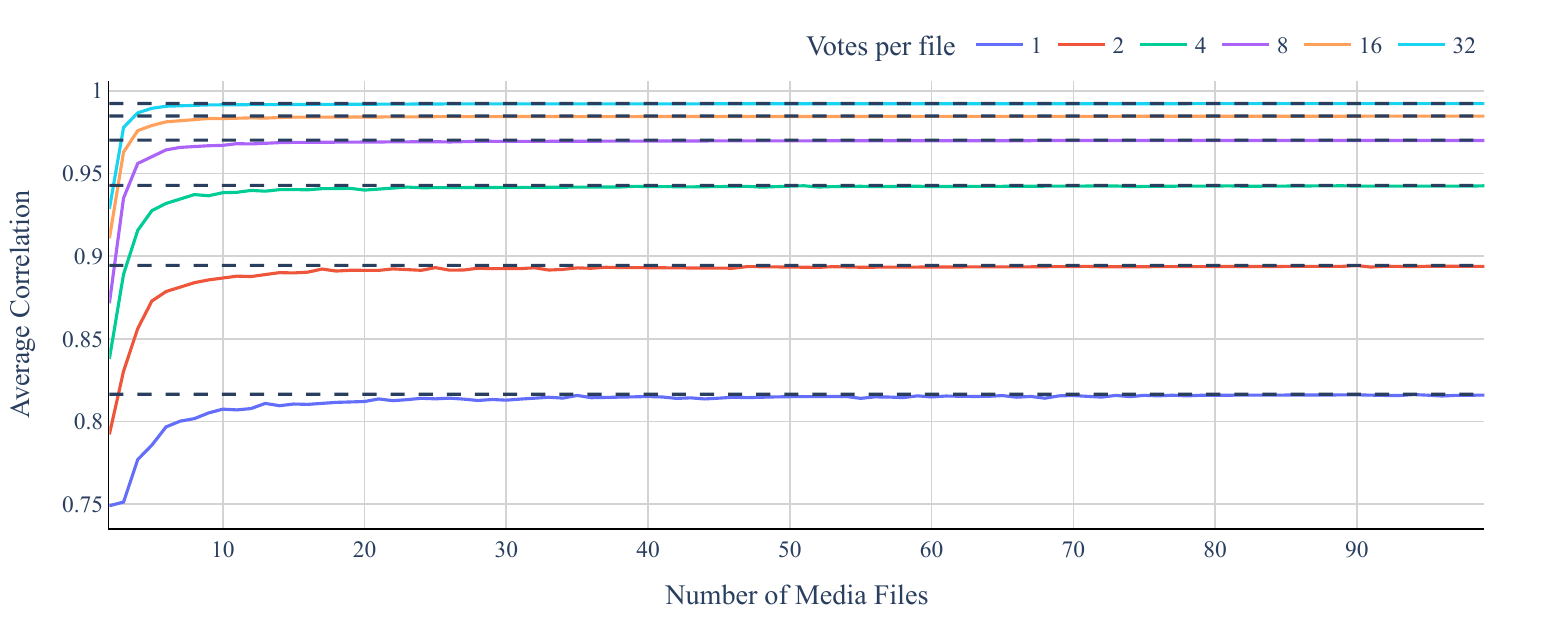}
    \caption{
    BinoVotes population correlation bounds from (\ref{eqn:BVPCCbound}) (dashed black)  and BinoVotes sample correlations by simulation (solid colors) vs number of files in sample.
    Sample correlations are averaged over 10,000 repetitions and this average rapidly approaches the population correlation bound. True quality distribution $f_Y$ is uniform.
    \vspace{0mm}
    }
    \label{fig:na-converge}
\end{figure}
For a given subjective test with $n_f$ media files and $n_f$ corresponding true quality values, PCC does depend on $n_f$.
The sample correlation, which is a random variable, captures this $n_f$ dependence while the population correlation (\ref{eqn:corr-pop-bound}) does not.
But Figure~\ref{fig:na-converge} shows that the expected value of the sample correlation converges rapidly to the population correlation as $n_f$ increases. 
This convergence is quicker for larger numbers of votes per file.
And regardless of the number votes per file, for any reasonably sized subjective test, say $50 \le n_f$, the difference between the sample correlation and the population correlation becomes insignificant. 
Thus we conclude that the population-based PCC bound that we have developed is quite satisfactory for all practical purposes.

\section{Application to Subjective Test Results}\label{sec:application}
To this point we have been focused primarily on modeling subjective votes for some given true quality.
Using only very basic assumptions about voting behavior, we derived general bounds for how well MOS can agree with true quality in terms of MSE and PCC in (\ref{eqn:mse-bound}) and (\ref{eqn:corr-bound2}), respectively.
We then developed the BinoVotes voting model and derived bounds relating BinoMOS results and true quality in (\ref{eqn:mse-bound-binovotes}) and (\ref{eqn:corr-pop-bound}).

We now shift our focus towards datasets and deriving performance bounds for arbitrary objective estimators on specific datasets in terms of MSE and PCC.\footnote{Software implementation of bounds and BinoVotes is available at \mbox{\url{https://github.com/NTIA/its-mos-agreement}}.}
We achieve this by again acknowledging that the best possible objective estimator is an oracle with direct access to true quality information.
Where previous results were presented in terms of the true quality distribution, we now focus on deriving results that only use information available from a sample of a MOS distribution.

For any subjective test results, the true quality distribution is hidden, and we only have access to a sample from the corresponding MOS distribution.
Thus, in order to compute bounds, we must first estimate the critical values needed in (\ref{eqn:mse-bound}) and (\ref{eqn:corr-bound2}): $\E{v_r(Y)}$ and $\var{X}$.
For convenience we label these, along with $\E{X}$, as
\begin{align}
    \mu_X = \E{X}, \sigma_X^2 = \var{X}, \sigma_V^2 = \E{v_r(Y)}.
\end{align}
We must estimate these values from data, and we label those estimates as $\hat{\mu}_X, \hat{\sigma}_X^2$, and $\hat{\sigma}_V^2$.
For a dataset with $n_v$ votes per file, the estimate for the MSE bound becomes
\begin{align}\label{eqn:mse-bound-est}
    \hat{d}^2
    &= \frac{\hat{\sigma}_V^2}{n_v},
\end{align}
and the estimate for the PCC bound becomes
\begin{align}\label{eqn:corr-bound-est}
    \hat{\rho} = \sqrt{\frac{\hat{\sigma}_X^2 - \frac{1}{n_v}\hat{\sigma}_V^2}{\hat{\sigma}_X^2}}.
\end{align}

We now consider two cases:
subjective test results with both MOS and vote variance for each file, $\{(x_i, \sigma_i^2)\}_{i=1}^{n_f}$, and subjective test results with only MOS values for each file, $\{x_i\}_{i=1}^{n_f}$.
In either case, the estimates of the mean and variance of the MOS distributions are the same:
\begin{align}\label{eqn:mos-stats}
    \hat{\mu}_X = \frac{1}{n_f}\sum_{i=1}^{n_f}x_i, \hspace{6mm}
    \hat{\sigma}_X^2 = \frac{1}{n_f - 1}\sum_{i=1}^{n_f} (x_i - \hat{\mu}_X)^2.
\end{align}
When vote variance information is provided for each file, then the estimate of the expected vote variance is just the average observed vote variance, which we define as
\begin{align}\label{eqn:average-vote-var}
    \hat{\sigma}^2_{DV} = \frac{1}{n_f} \sum_{i=1}^{n_f} \sigma_i^2.
\end{align}
When vote variance information is available, the estimates provided by (\ref{eqn:mos-stats}) and (\ref{eqn:average-vote-var}) are used to calculate the bounds in (\ref{eqn:mse-bound-est}) and (\ref{eqn:corr-bound-est}).
These results are the fully data-driven performance bounds.

When vote variance information is not available, we must make an assumption to generate a value of $\hat{\sigma}_V^2$.
One solution would be to use an empirical result from other tests that do provide vote variance information.
A natural candidate would be to take the global average value of all available values of $\hat{\sigma}_{DV}^2$ and use it as a fixed value, which we denote as the global average observed vote variance, $\hat{\sigma}_{GV}^2$.
In this case, $\hat{\sigma}_{GV}^2$ would be used as $\hat{\sigma}_V^2$ in (\ref{eqn:mse-bound-est}) and (\ref{eqn:corr-bound-est}).

Another option would be to use the BinoVotes voting model and the observed MOS sample to estimate the average vote variance.
We define $\hat{\sigma}_{BV}^2$ as the average vote variance under a BinoVotes model.
From (\ref{eqn:bv-expected-vote-varianccc}) and (\ref{eqn:mos-expected-value}) we see
\begin{align}\label{eqn:bv_var}
    \hat{\sigma}_{BV}^2 &= \frac{(\hat{\mu}_X - s_L)(s_H - \hat{\mu}_X) - \hat{\sigma}_{Y}^2}{n_s - 1}.
\end{align}
By rearranging (\ref{eqn:mos-variance}) and using the BinoVotes vote variance estimate we can see that
\begin{align}\label{eqn:quality-variance-est}
    \hat{\sigma}_Y^2 = \hat{\sigma}_X^2 - \frac{\hat{\sigma}_{BV}^2}{n_v}.
\end{align}
By substituting (\ref{eqn:quality-variance-est}) into (\ref{eqn:bv_var}), and solving for $\hat{\sigma}_{BV}^2$, we find the expected vote variance under BinoVotes for a given a sample from a MOS distribution to be
\begin{align}\label{eqn:binovotes-vote-var-est}
    \hat{\sigma}_{BV}^2 = \frac{n_v}{n_m - 1}\left((\hat{\mu}_X - s_L)(s_H - \hat{\mu}_X) - \hat{\sigma}_X^2 \right).
\end{align}
In this case $\hat{\sigma}_{BV}^2$ would be used as $\hat{\sigma}_V^2$ in (\ref{eqn:mse-bound-est}) and (\ref{eqn:corr-bound-est}).

\section{Analysis of 18 Subjective Test Results}\label{sec:data-analysis}
We now calculate RMSE and PCC bounds using subjective test results from 2 image subjective tests, 2 video subjective tests, and 14 speech subjective tests.
We selected these tests as examples of those that include vote variance information along with MOS values. 
Vote variance information can be provided in the form of raw votes or a variance or standard deviation value that accompanies each MOS value.
We note that many subjective test results are shared without this information and that including it would add significant value.

Vote variance information allows us to calculate the fully data-driven performance bounds.
We compare these with the two alternatives presented in Section \ref{sec:application}.
The first alternative is to calculate the bounds using the 
global average of the observed vote variances $\hat{\sigma}_{GV}^2$.
The second alternative is to calculate the bounds using  
the vote variance $\hat{\sigma}_{BV}^2$ obtained from the BinoVotes model and observed MOS sample using (\ref{eqn:binovotes-vote-var-est}).
These comparisons show how bounds change when we are forced to compensate for missing vote variance information.

\begin{figure}[t]
    \centering
    \includegraphics[width=\linewidth]{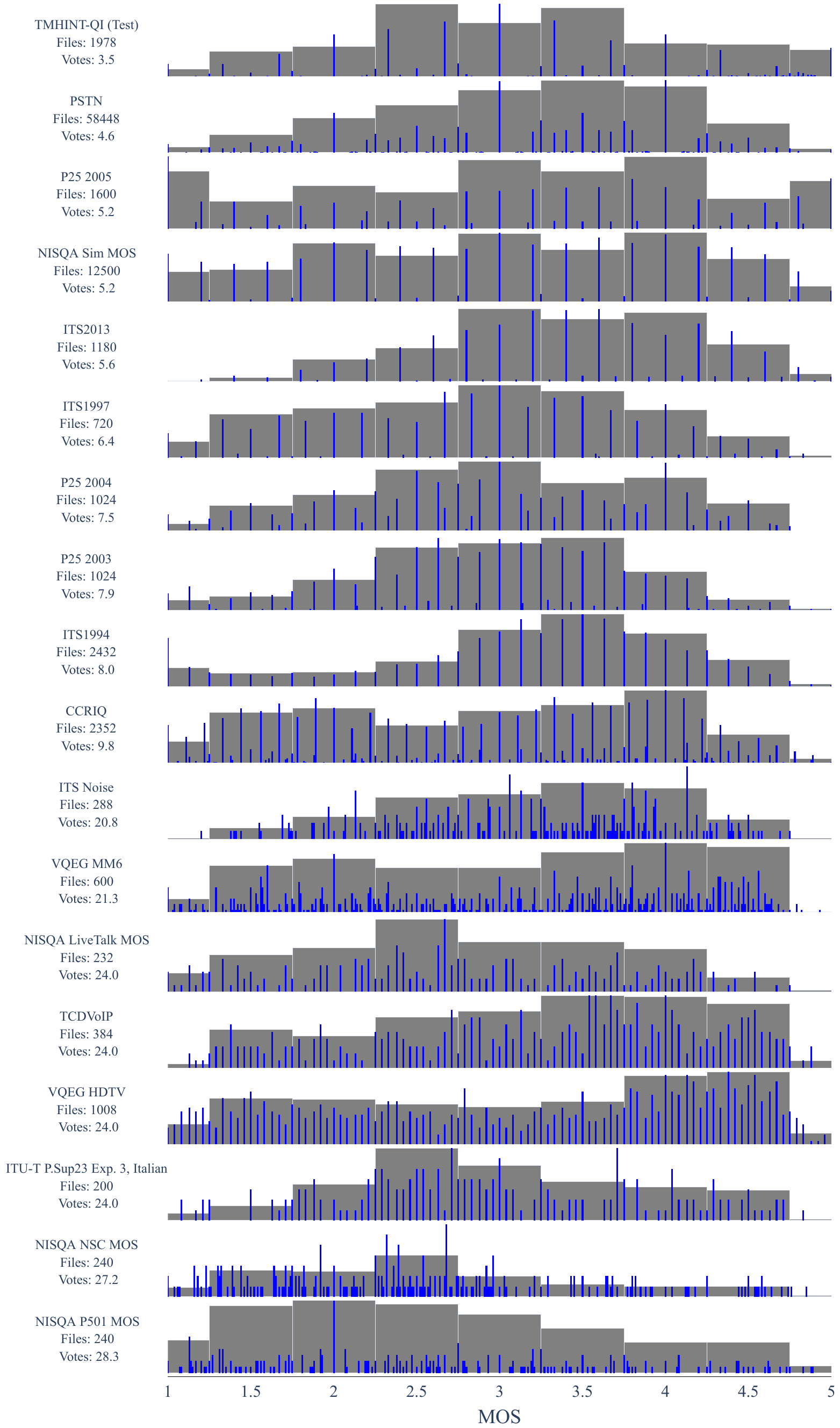}
    \caption{MOS distributions for 18 subjective tests. Low-resolution gray histograms emphasize distribution shape and high-resolution blue histograms emphasize distribution resolution, as determined by number of votes per file. \vspace{0mm}}
    \label{fig:individual-vote-MOS-dists}
\end{figure}
\begin{table}[t]
    \centering
    \caption{
    Subjective tests that include variance information ordered by average number of votes per file.
    \textdagger indicates image quality tests;
    \textdagger\textdagger indicates video quality tests; all others are speech quality tests.
    *Indicates tests of the land-mobile radio system in
    \cite{p25-subjective-tests-apco}.
    }
    \resizebox{\linewidth}{!}{\begin{tabular}{|l|l|l|l|l|l|} \hline 
	Name & $n_v$ & $\hat{\mu}_X$ & $\hat{\sigma}^2_X$ & $\hat{\sigma}_V^2$ \\ \hline 
	TMHINT-QI (Test) \cite{TMHINT} & 3.52 & 3.11 & 0.99 & 0.93 \\ \hline 
	PSTN \cite{Mittag2020Interspeech} & 4.57 & 3.12 & 0.76 & 0.80 \\ \hline 
	P25 2005* & 5.20 & 3.05 & 1.51 & 0.48 \\ \hline 
	NISQA Sim MOS \cite{Mittag2021IS} & 5.24 & 2.99 & 1.20 & 0.54 \\ \hline 
	ITS2013 \cite{VoranICASSP2013} & 5.64 & 3.41 & 0.60 & 0.83 \\ \hline 
	ITS1997 (Tests 16 and 17 in \cite{VoranSCW99}) & 6.40 & 2.85 & 0.89 & 0.58 \\ \hline 
	P25 2004* & 7.47 & 2.95 & 0.87 & 0.73 \\ \hline 
	P25 2003* & 7.91 & 2.92 & 0.65 & 0.74 \\ \hline 
	ITS1994 (Test 4 in \cite{MNBpartII}) & 8.00 & 3.12 & 0.88 & 0.63 \\ \hline 
	CCRIQ \cite{CCRIQ} \textdagger & 9.85 & 2.89 & 1.10 & 0.63 \\ \hline 
	ITS Noise \cite{ITSimageNoise} \textdagger & 20.84 & 3.18 & 0.63 & 0.69 \\ \hline 
	VQEG MM6 \cite{PinsonMM6} \textdagger\textdagger & 21.30 & 3.06 & 1.15 & 0.66 \\ \hline 
	NISQA LiveTalk MOS \cite{Mittag2021IS} & 24.00 & 2.76 & 0.89 & 0.66 \\ \hline 
	TCDVoIP \cite{Harte2015} & 24.00 & 3.22 & 0.99 & 0.62 \\ \hline 
	VQEG HDTV \cite{VQEGHDTVreport} \textdagger\textdagger & 24.00 & 3.10 & 1.33 & 0.51 \\ \hline 
	ITU-T P.Sup23 Exp. 2, Italian \cite{Psup23} & 24.00 & 2.95 & 0.80 & 0.57 \\ \hline 
	NISQA NSC MOS \cite{Mittag2021IS} & 27.17 & 2.55 & 0.91 & 0.48 \\ \hline 
	NISQA P501 MOS \cite{Mittag2021IS} & 28.33 & 2.60 & 1.04 & 0.42 \\ \hline 
\end{tabular}}
    \label{tab:individual-vote-datasets}
\end{table}
Table~\ref{tab:individual-vote-datasets} and Figure~\ref{fig:individual-vote-MOS-dists} summarize the 18 subjective tests.
The number of files per test ($n_f$) ranges from 200 to 58,448 and the average number of votes per file ($n_v$) ranges from 3.52 to 28.33. The MOS means range from 2.55 to 3.41 and MOS variances range from 0.63 to 1.51. The mean vote variance ranges from 0.42 to 0.93.
Figure~\ref{fig:individual-vote-MOS-dists} shows the MOS distributions.
We only used tests that cover the full MOS range from 1 to 5.
More specifically, we created 8 uniform bins of width 0.5 from 1 to 5, and included only subjective tests where all 8 bins were non-empty.

\begin{figure}[h]
    \centering
    \subfloat[]{
        \includegraphics[width=0.95\linewidth]{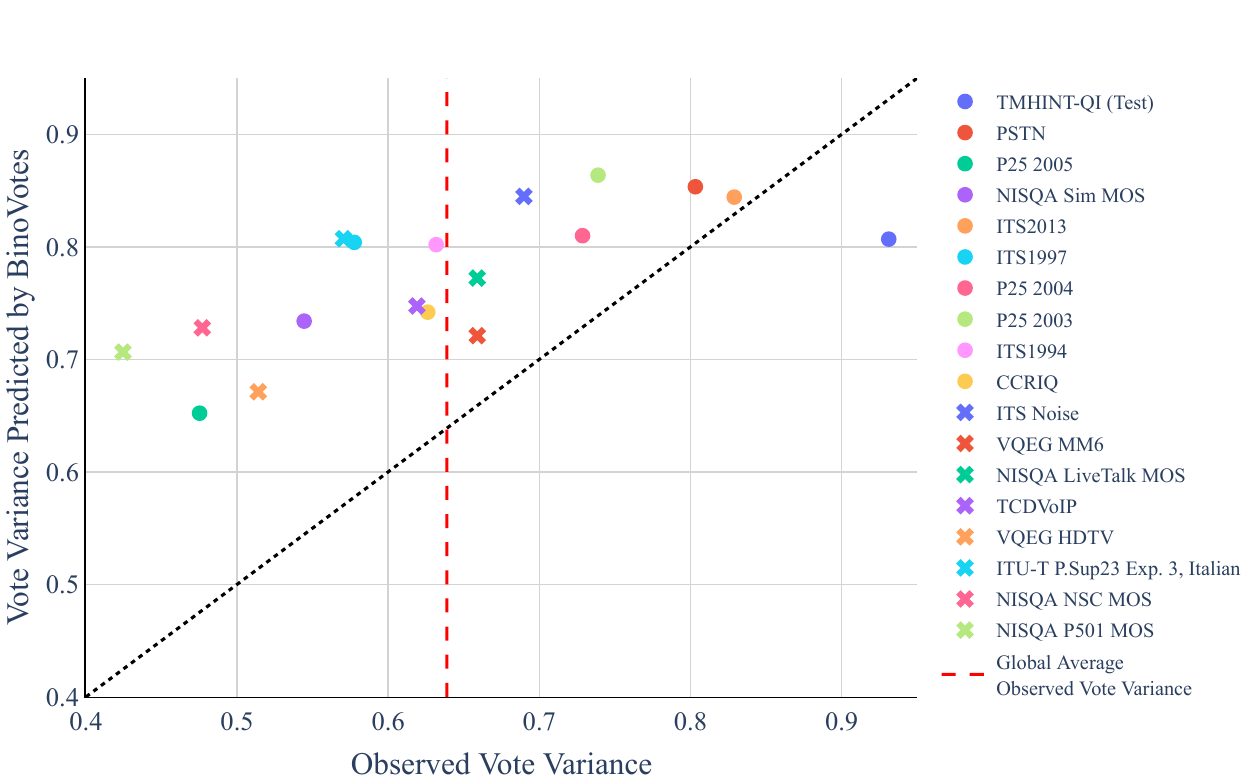}
        \label{fig:expected-vote-variance-comp}
        }
        \\ \vspace{-4mm}
    \subfloat[]{
        \includegraphics[width=0.95\linewidth]{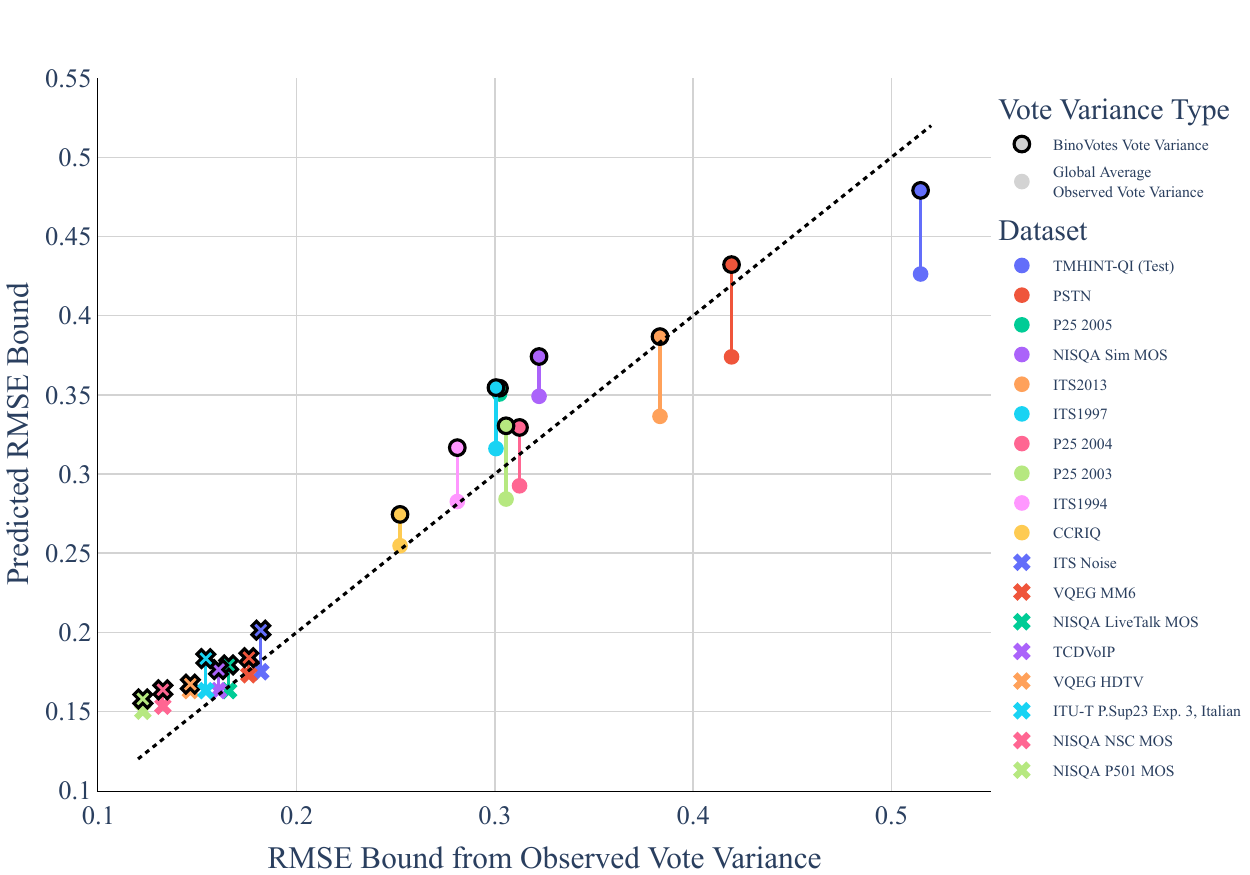}
        \label{fig:rmse-comp}
        }
        \\ \vspace{-4mm}
    \subfloat[]{
        \includegraphics[width=0.95\linewidth]{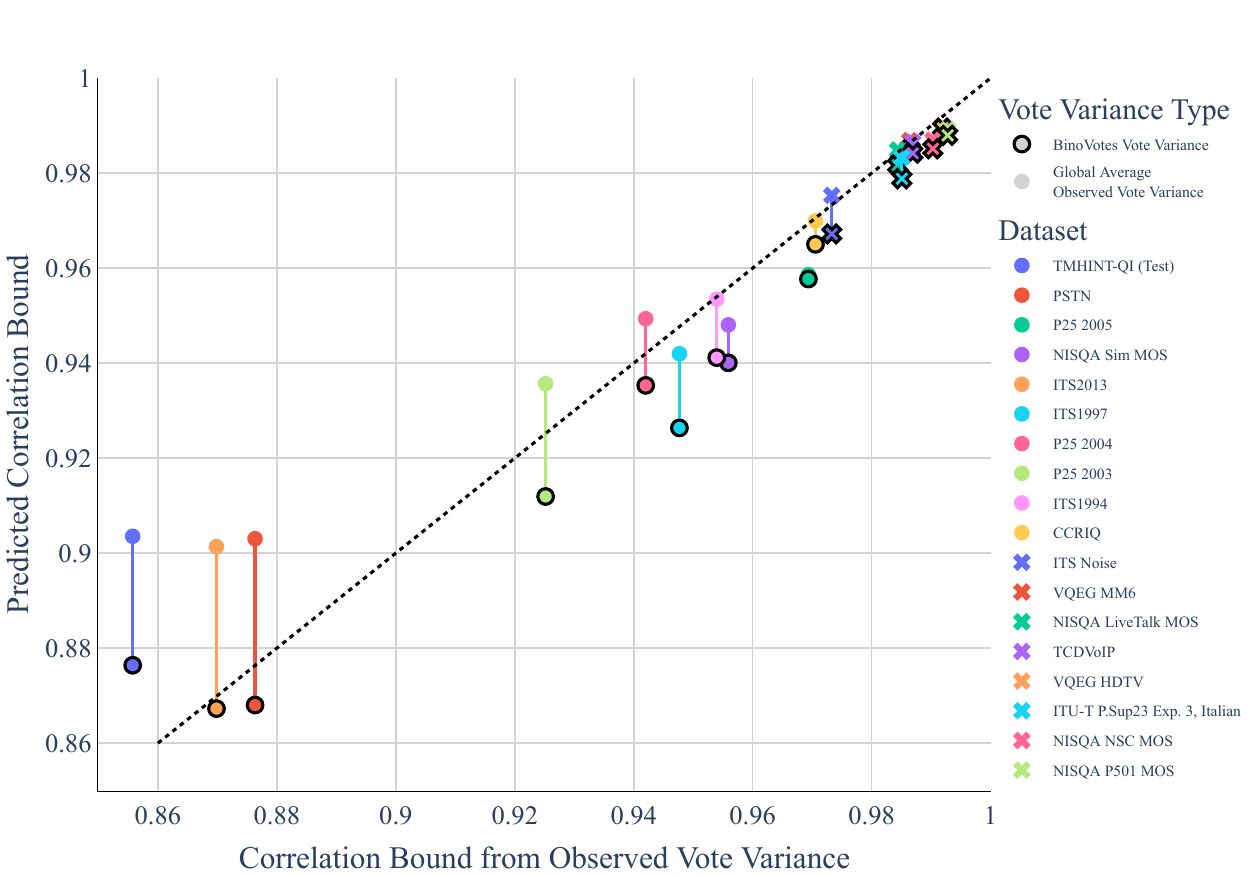}
        \label{fig:corr-comp}
    }
    \caption{BinoMOS results compared to results from 18 subjective tests. (a) Expected vote variance under BinoVotes compared to the average observed vote variance for each test. (b) RMSE lower bounds under BinoVotes and under global average observed vote variance versus fully data-driven bounds. (c) PCC upper bounds under BinoVotes and under global average observed vote variance versus fully data-driven bounds.  \vspace{0mm}}
    \label{fig:vote-variance-comparisons}
\end{figure}

\subsection{Agreement Statistic Bound Comparisons}
The performance bounds in (\ref{eqn:mse-bound-est}) and (\ref{eqn:corr-bound-est}) depend on the MOS distribution variance estimate $\hat{\sigma}_X^2$, the expected vote variance estimate $\hat{\sigma}_V^2$, and the number of votes per file $n_v$.
The estimate $\hat{\sigma}_X^2$ and value $n_v$ are the same for all three methods of computing bounds given in Section~\ref{sec:application}.
The expected vote variance $\hat{\sigma}_V^2$ is the only difference, so we compare those values first.

Figure~\ref{fig:expected-vote-variance-comp} shows the observed average vote variance $\hat{\sigma}_{DV}^2$ and the vote variance from the BinoVotes model, $\hat{\sigma}_{BV}^2$.
It also shows the global average observed vote variance, which is $\hat{\sigma}_{GV}^2=0.64$ for these 18 tests.
The standard deviation of observed vote variance values across the tests is 0.13.
The minimum observed average vote variance is 0.42 from NISQA P501 MOS and the maximum is 0.93 from TMHINT-QI (Test). The figure shows that for 17 of the 18 tests (all but TMHINT-QI), BinoVotes produces vote variances that are higher than those observed in the data.
One possible explanation for this behavior is that during a test, subjects may reserve the extremes of the rating scale for media files with the most extreme attributes, whereas BinoVotes does not.
In other words, most subjects will know that something of middling quality is neither bad nor excellent, and will be less likely than BinoVotes to cast such a vote. 
The largest absolute difference is 0.28 for NISQA P501 MOS, where BinoVotes predicts an expected vote variance of 0.70.
The average difference between BinoVotes' predicted average vote variance and observed variance is 0.13.

In spite of these differences in vote variance, the BinoVotes model provides very useful RMSE and PCC bounds. 
Figure~\ref{fig:rmse-comp} shows the lower bounds for RMSE when using $\sigma_{GV}^2$ or $\sigma_{BV}^2$ ($y$-axis) compared to the fully data-driven $\sigma_{DV}^2$ ($x$-axis). BinoVotes-based bounds are always larger than the global average-based bounds, and they also tend to be larger than the fully data-driven bounds, with TMHINT-QI (Test) being the sole exception again.
The maximum observed difference between the global average-based bounds and the fully data-driven bounds is 0.09 for TMHINT-QI (Test).
This is directly attributable to this subjective test having the largest observed vote variance which is also the furthest from the global average.
The average difference between the global average-based bounds and the fully data-driven bounds is $-$0.004, demonstrating that, while the predicted bounds are not perfect, they are still informative.
The maximum observed difference between the BinoVotes-based bounds and the fully data-driven bounds is 0.05 for ITS1997 and the average observed difference is 0.02.
The RMSE bound values in this work range from 0.12 to 0.51, and all of the differences noted here are modest compared to these values and the range that they cover.

Figure~\ref{fig:corr-comp} shows the same comparisons as \ref{fig:rmse-comp} but for PCC upper bounds and similar patterns emerge.
BinoVotes-based bounds are always lower than global-average-based bounds, and BinoVotes tends to underpredict the correlation bound while the global average-based bounds sometimes overpredict and sometimes underpredict (as expected due to the nature of utilizing the global average observed vote variance).
The maximum observed difference for the global-average-based bounds is 0.05, again for TMHINT-QI (Test), while the average difference is 0.005.
The maximum observed difference for the BinoVotes bounds is 0.021, again for ITS1997, while the average difference is \mbox{$-$0.006}.
For context, the range of PCC values covers 0.86 to 0.99.
So while the estimated bounds are not perfect, they clearly provide valuable information in cases where we do not have vote variance information.

\subsection{Subjective Tests without Variance Information}
Subjective test results are sometimes distributed without any variance information --- they include just MOS values and the number of votes per file. 
Table \ref{tab:realDataBounds} lists four such subjective tests that are commonly used in research advancing speech quality estimation.
For these tests we can use both the global average observed vote variance and the BinoVotes model to find PCC (\ref{eqn:BVPCCbound}) and MSE (\ref{eqn:BVMSEbound}) bounds.
Note that BinoVotes is flexible to the rating scale used but the global average observed vote variance $\sigma_{GV}^2$  must be calculated over data with a single rating scale and can be applied only to data with that same scale.
Since the IU data is not on the standard 1 to 5 MOS scale we 
do not provide bound estimates using $\hat{\sigma}_{GV}^2$ for IU in Table~\ref{tab:realDataBounds}.




\begin{table}[t]
    \centering
    \caption{Four subjective tests that do not include vote variance information.  Bounds can still be calculated using BinoVotes or by using a fixed value of $\hat{\sigma}_{GV}^2 = 0.64$.}
    \resizebox{\linewidth}{!}{\begin{tabular}{|l|l|l|l|l|l|l|l|l|} \hline 
	\multirow{2}{*}{Dataset} & \multirow{2}{*}{$\hat{\mu}_X$} & \multirow{2}{*}{$\hat{\sigma}^2_X$} & \multirow{2}{*}{$n_v$} & \multicolumn{2}{l|}{RMSE Bound} & \multicolumn{2}{l|}{PCC Bound} \\ \cline{5-8} 
 &  &  &  & BinoVotes & Fixed & BinoVotes & Fixed \\ \hline
	VCC18 & 2.92 & 0.79 & 4 & 0.46 & 0.40 & 0.85 & 0.89 \\ \hline 
	IU & 5.25 & 4.56 & 5 & 0.64 & NA & 0.95 & NA \\ \hline 
	VMC22 & 2.93 & 0.85 & 8 & 0.32 & 0.28 & 0.94 & 0.95 \\ \hline 
	Tencent & 2.85 & 1.38 & 20 & 0.18 & 0.18 & 0.99 & 0.99 \\ \hline 
\end{tabular}}
    \label{tab:realDataBounds} \vspace{-4mm}
\end{table}

\section{Conclusion}

We have derived bounds for the agreement between a set of subjective test results and an objective estimator of those results.  
More specifically, we have derived an upper bound on PCC and a lower bound on MSE.
The only assumption required is that for any media file tested, the subjective test responses are drawn from a distribution that has an expected value that is the true value for that media file.

These results are highly actionable. 
For any given subjective test, one can use (\ref{eqn:mse-bound-est}) and a set of simple calculations to find a lower bound on the MSE that any objective estimator can obtain on that test. 
Similarly, one can use (\ref{eqn:corr-bound-est}) and a set of simple calculations to find an upper bound on the PCC that any objective estimator can obtain on that test. 
These bounds have great value because for each dataset they provide appropriate goals for researchers who are designing objective estimators. 
It is natural, but not attainable, for researchers to strive for an MSE value of zero and a PCC value of one. 
The bounds given in this paper provide realistic goals that are appropriate for each individual dataset.

For datasets that contain vote variance information, the bounds are fully data-driven and depend only on the observed average vote variance and the variance of the MOS distribution.
Note that the observed average vote variance includes the effects of per-subject biases --- greater biases increase the observed average vote variance and this in turn reduces the PCC upper bound and increases the MSE lower bound.

For datasets without vote variance information we can assume votes come from the binomial-inspired  BinoVotes process and then estimate the bounds.
We have considered the results of 18 subjective speech and video tests that provide vote variance information, covering a total of 86,450 files and over 493,000 votes. 
We found that BinoVotes-based bounds are in general accurate and informative and provide a foothold for bounding performance for datasets that do not provide vote variance information.
This is in spite of the fact that BinoVotes often produces vote variances that are somewhat higher than those observed in the data. 
Adding per subject-biases to the BinoVotes model would increase the variance of the votes it produces and would thus broaden the gap between the model output and the observed data.  
Since the vote variance drives the bounds, the bounds would also move away from what is most correct.
Also, in any situation where the over-estimation of vote variance by BinoVotes could be problematic, one can instead choose to use the global average observed vote variance, $\hat{\sigma}_{GV}^2$ provided here.


Note that these bounds do not apply to seen training data while training a machine learning-based estimator --- of course an estimator can learn any and all noise that is in training data.  But when using an estimator on unseen, non-training data the bounds do apply.
And these bounds are in fact expected values, so it is possible for an objective estimator to outperform them on occasion. But it should not be possible to consistently outperform them.

We have applied the results developed here to 22 different subjective tests which is significant but not exhaustive. 
A potential future effort could include additional datasets, and could focus on video, image, and multimedia areas which are admittedly underrepresented here. 
An additional candidate for future work would be extending results presented here to address other agreement statistics of interest to the research community. 
A top candidate for this effort would be SRCC.



\bibliographystyle{IEEEtranDOI}
\bibliography{sources-clean}

\end{document}